\documentclass[useAMS,usenatbib]{mn2e}
\usepackage{graphicx}
\title{Indirect evidence of GeV Dark Matter}
\author[Chan]{Man Ho Chan
\thanks{mhchan@phy.cuhk.edu.hk}\\
Department of Physics, The Chinese University of Hong Kong \\
Shatin, New Territories, Hong Kong, China}

\begin{document}

\date{Accepted XXXX, Received XXXX}

\pagerange{\pageref{firstpage}--\pageref{lastpage}} \pubyear{XXXX}

\maketitle

\label{firstpage}

\begin{abstract}
Recently, an excess of GeV gamma ray near the Galactic Centre has been 
reported. The spectrum obtained can be best fitted with the annihilation 
of $30-40$ GeV dark matter particles through $b \bar{b}$ channel. In this 
letter, I show that this annihilation model can also solve the mysteries 
of heating source in x-ray plasma and the unexpected high gamma-ray 
luminosity. The cross section constrained by these observations 
give excellent agreements with both the predicted range by using Fermi-LAT 
data and the canonical thermal relic abundance cross section.
\end{abstract}


\section{Introduction}
Recently, high energy gamma ray observations reveal some excess emissions 
near the Galactic Centre. These excess emissions cannot be easily 
explained by standard physical processes. One potential origin of such 
emissions is due to an unusual population of millisecond pulsars 
\citep{Gordon,Abazajian}. However, \citet{Daylan} point out that the 
large diffuse signal of gamma ray disfavours the possibility of pulsar 
emissions. Even including both known sources and unidentified sources, 
the millisecond pulsars can only account no more than 10 percent of the 
GeV excess \citep{Hooper}. In fact, the majority of discussions of the GeV 
excess is now focused on the annihilation 
of dark matter particles \citep{Gordon,Abazajian,Daylan,Izaguirre,Calore}. 
It is also because the 
gamma ray spectrum obtained from Fermi-LAT data can be well fitted with 
$b \bar{b}$ annihilation channel of dark matter particles 
\citep{Abazajian,Daylan}. The required cross section and rest mass of 
dark matter particle are $<\sigma v>=(1.4-7.5) \times 10^{-26}$ cm$^3$ 
s$^{-1}$ and 
$m_{\chi} \sim 30-40$ GeV respectively \citep{Abazajian,Daylan}. This 
cross section is consistent with the 
expected canonical thermal relic abundance cross section 
($<\sigma v> \approx 3 \times 10^{-26}$ cm$^3$ s$^{-1}$) in cosmology. 
Furthermore, the inner slope of the radial-depedence of the gamma ray 
emissions is $\gamma \approx 1.1-1.3$ (the best fit is $\gamma=1.26$) 
\citep{Daylan}, which is 
consistent with the numerical simulation of dark matter halo structure 
$\gamma=1-1.5$ \citep{Navarro,Moore}.

In this letter, I 
show that the $b \bar{b}$ annihilation channel can also explain the energy 
required in x-ray emissions in the Galactic Centre. This evidence can 
further support the dark matter annihilation model and constrain 
the cross section and rest mass of the dark matter particles.

\section{X-ray emission at Galactic Centre}
In the past decade, a large amount of diffuse x-ray data had been 
obtained by {\it Chandra, BeppoSAX, Suzaku} and {\it XMM-Newton} 
\citep{Sidoli,Muno,Sakano,Uchiyama}. In particular, \citet{Muno} use the 
data from {\it Chandra} to model the temperature of the two components 
within 20 pc as 0.8 keV (soft component) and 8 keV (hard component). The 
energy required to sustain the 0.8 keV 
and 8 keV components are $3 \times 10^{36}$ erg/s and $10^{40}$ erg/s 
respectively \citep{Muno}. Later, \citet{Belmont} point out that the 
cooling by 
adiabatic expansion may not be important if the hard component could 
actually be a gravitationally confined helium plasma. Therefore, the 
actual energy required for the hard component to balance the radiative 
cooling would be $(1.4-2.6) \times 10^{36}$ erg/s 
within 20 pc \citep{Muno}, but not $10^{40}$ erg/s. Although supernova 
explosions are able to 
provide such a high energy, it is not possible for supernovae to heat 
the hard component plasma to such a high temperature \citep{Uchiyama}. 
Also, the 
correlation between the hard and soft emission suggests that they are 
produced by related physical processes \citep{Muno}. Therefore, the 
required energy to balance the radiative coolings of both soft and hard 
component might be given by some other origins. However, there is no 
widely accepted mechanism to heat and sustain the plasma to such a high 
temperautre \citep{Muno,Uchiyama}. One potential explanation is that the 
heating might 
result from the viscous friction on molecular clouds flowing toward the 
Galactic Centre \citep{Belmont}. 

Here, I propose that the energy given out by annihilation of dark matter 
particles can explain the energy requirement of both soft and hard 
components ($(4.4-5.6) \times 10^{36}$ erg/s) within 20 pc. Although 
a large amount of energy from annihilation is given out in the form of 
gamma ray which is nearly transparent to the plasma, a large amount of 
high energy electrons and positrons can also be produced 
through the $b \bar{b}$ annihilation channel. The spectrum of positron (or 
electron)
energy $dN_e/dE$ per one annihilation is shown in Fig.~1 
\citep{Borriello,Crocker}. These high energy 
electrons and positrons would lose their energy and give their 
energy to the plasma mainly by three different processes: ionization 
loss $\dot{E}_{ion}$, synchrotron loss $\dot{E}_{syn}$ and inverse Compton 
scattering $\dot{E}_{IC}$. The corresponding energy loss rates 
are \citep{Longair}:
\begin{equation}
\dot{E}_{ion}=7.64 \times 10^{-9} \left( \frac{n_e}{1~{\rm cm^{-3}}} 
\right) \left(3 \ln \frac{E}{m_ec^2}+19.8 \right)~{\rm eV/s},
\end{equation}
\begin{equation}
\dot{E}_{syn}=6.6 \times 10^{-10} \left(\frac{E}{m_ec^2} \right)^2 \left( 
\frac{B}{10^{-3}~\rm G} \right)^2~{\rm eV/s},
\end{equation}
and
\begin{equation}
\dot{E}_{IC}=1.6 \times 10^{-9} \left(\frac{E}{m_ec^2} \right)^2 
\left(\frac{U_{rad}}{6 \times 10^4~\rm eV/cm^3} \right)~{\rm eV/s},
\end{equation}
where $U_{rad}$ is the radiation energy density, $n_e$ and $B$ are the 
number density of electrons and magnetic 
field strength in the plasma respectively. The total number and total 
energy of positrons or electrons produced per one annihilation are 
$N_e=\int (dN_e/dE)dE \approx 12$ and 
$\bar{E}=\int E(dN_e/dE)dE \approx 6-8$ GeV respectively for 
$m_{\chi}=30-40$ GeV. Therefore, the average energy for one positron or 
electron produced is $E \approx 0.5-0.7$ GeV, which gives $E/m_ec^2 \sim 
10^3$. Since $n_e \sim 0.1$ cm$^{-3}$ \citep{Muno}, $U_{rad} \sim 10^4$ 
eV/cm$^3$ \citep{Wolfire,Fritz} and $B \sim 
10^{-4}-10^{-3}$ G in the plasma near the Galactic centre \citep{Crocker}, 
the total energy loss rate is 
$\dot{E}= \dot{E}_{ion}+\dot{E}_{syn}+\dot{E}_{IC} 
\sim 10^{-7}-10^{-3}$ eV/s. The cooling rate would first be dominated by 
inverse Compton Scattering and synchrotron loss. When the positron or 
electron is cooled down to about 1 MeV, the cooling rate would 
be dominated by ionization loss. As a result, the required cooling time is 
$t_c \sim 10^{13}-10^{14}$ s (see Fig.~2). 

For the diffusion process of the positrons or electrons, let's first 
consider the simple random walk model. The stopping distance of 
a high energy positron or electron is $d_s \sim \sqrt{r_L \times ct_c}$ 
\citep{Boehm}, 
where $r_L=E/eB$ is the Larmor radius. For $B \sim 10^{-3}$ G, we 
have $d_s<1$ pc, which is very small compared with the size of our 
interested region (20 pc). Therefore, the required cooling time is short 
enough such that the diffusion process is not important. Nevertheless, 
\citet{Regis} suggest that the amplitude of the random magnetic field and 
the turbulence effect are also important to the diffusion process. 
This kind of diffusion can be described by a diffusion coefficient $K_0$ 
and an index $\delta$. For a large scale (greater than 100 pc), the ranges 
of the values are $K_0=10^{27}-10^{30}$ cm$^2$ s$^{-1}$ and 
$\delta=0.3-0.6$ \citep{Delahaye,Regis,Lacroix}. For the innermost region 
near the 
Galactic Centre, the picture is much more uncertain. \citet{Regis} reveal 
from the analysis of $\gamma$-ray observations that a significant 
reduction of the diffusion coefficient in the inner 10 pc region is found. 
They apply two models, namely Kraichnan and Kolmogorov, to obtain the 
characteristic diffusion length $d_f$. They get $d_f \sim 10$ pc and $d_f 
\sim 30$ pc by using the Kraichnan model and Kolmogorov model 
respectively. 
Therefore, not all of the energy of positrons or electrons is lost due to 
the cooling process during the diffusion within $d_f$. The diffusion 
length for an electron with initial energy $E_i$ is given by 
\citep{Fornengo}
\begin{equation}
d_f= \left[4 \int_{E_f}^{E_i} \frac{K_0E^{\delta}}{\dot{E}}dE 
\right]^{1/2},
\end{equation}
where $E_f$ is the energy of the electron after the diffusion of length 
$d_f$. Let's consider the lower bounds of the parameters from 
large scale diffusion $K_0=10^{27}$ 
cm$^2$ s$^{-1}$ and $\delta=0.3$. For $E_i=0.6$ GeV and $d_f=20$ 
pc, we get $E_f=0.2$ GeV, which means over 65\% of the energy would be 
lost during the diffusion process.  

In the following discussion, we first neglect the effect of diffusion. The 
result would not be affected if the simple random walk model is the 
correct diffusion model. However, if either of the other two 
turbulence models is the correct diffusion model, the calculated cross 
section would be at most increased by a factor of 1.5.

\begin{figure}
 \includegraphics[width=80mm]{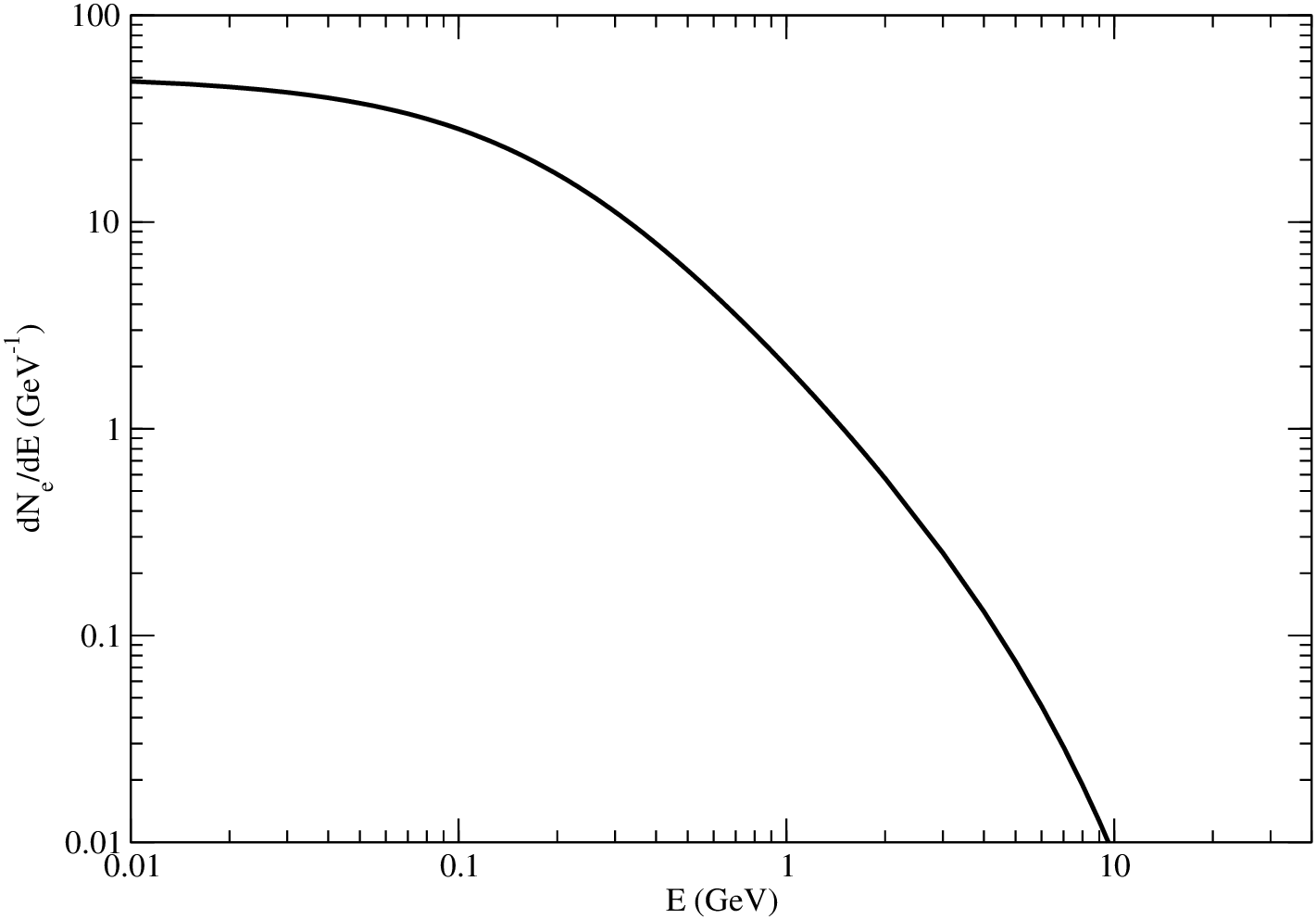}
 \caption{The energy spectrum of positron produced per one annihilation 
\citep{Borriello}.}
\vskip5mm
\end{figure}

The total annihilation rate within $R=20$ pc is given by
\begin{equation}
\Psi (R)=\int_0^R \rho^2 <\sigma v>m_{\chi}^{-2} 4 \pi r^2dr,
\end{equation}
Here, $\rho$ is the dark matter density profile, which is assumed to be 
the generalized NFW profile \citep{Cirelli}:
\begin{equation}
\rho=\rho_{\odot} \left( \frac{r}{r_{\odot}} \right)^{-\gamma} \left[ 
\frac{1+(r/r_s)}{1+(r_{\odot}/r_s)} \right]^{-3+\gamma},
\end{equation}
where $\rho_{\odot}=0.3$ GeV/cm$^3$, $r_{\odot}=8.5$ kpc and $r_s=20$ kpc. 
In our calculations, we assume $\gamma=1.26$, which is the best fit of 
the observed gamma ray spectrum by Fermi-LAT \citep{Daylan}. The total 
energy loss due to electron-positron pairs is $\approx 2 \Psi (R) \times 
\bar{E}$. Since the energy required 
to sustain the soft and hard components is $(4.4-5.6) \times 10^{36}$ 
erg/s, we can constrain the parameter space of $m_{\chi}$ and $<\sigma v>$ 
by using Eq.~(5) (see Fig.~3). 
In the plot, we see that the allowed parameter space falls within the 
range predicted by the annihilation model from Fermi-LAT data $<\sigma 
v>=(1.4-7.5) \times 10^{-26}$ cm$^3$ s$^{-1}$ for $m_{\chi}=30-40$ GeV.

\begin{figure}
 \includegraphics[width=80mm]{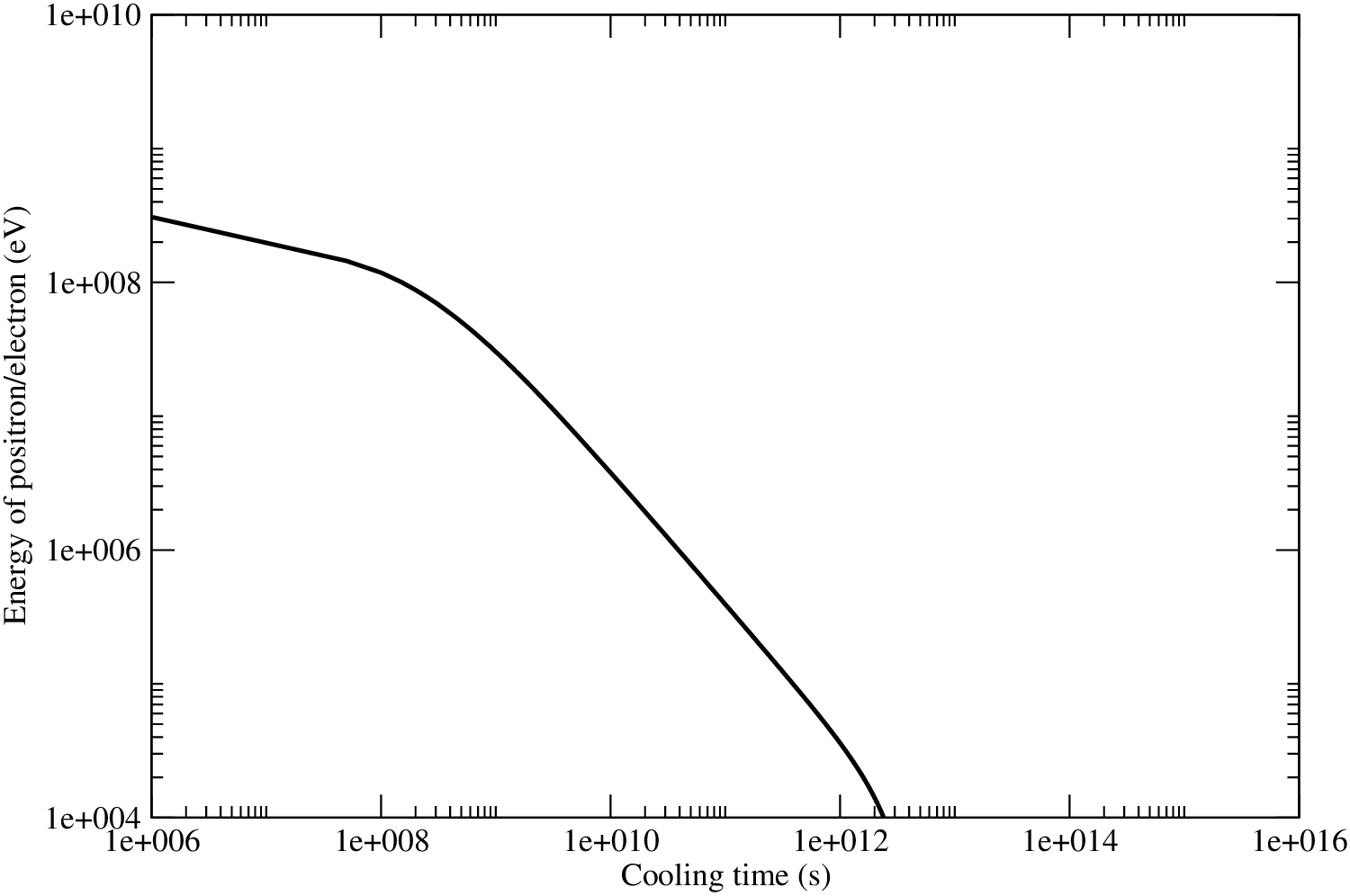}
 \caption{The energy of a positron or electron as a function of time 
during cooling process for $B=10^{-3}$ G and $U_{rad}=6 
\times 10^4$ eV/cm$^3$. We have assumed $n_e=0.1$ cm$^{-3}$ and $E=0.6$ 
GeV initially.} 
\end{figure}

For consistency checking, the above result also agrees with the total 
luminosity of gamma ray with energy greater than 500 MeV within $R'=30$ pc 
in the Galactic Centre detected 
by the EGRET telescope \citep{Mayer,Cheng}. The total energy released per 
one annihilation is given by:
\begin{equation}
\dot{E}=2m_{\chi}^{-1}c^2 \int_0^{R'}\rho^2<\sigma v> 4 \pi r^2dr.
\end{equation}
Since the energy carried by neutrinos is negligible \citep{Bergstrom}, 
by using the predicted range of $<\sigma v> \approx (2-4)\times 10^{-26}$ 
cm$^3$ s$^{-1}$ from x-ray emission and $m_{\chi}=30-40$ GeV, the total 
luminosity of gamma ray is 
$\dot{E}_{\gamma} \approx \dot{E}-2 \Psi (R') \bar{E} \approx (2.0-2.7) 
\times 10^{37}$ erg/s, which agrees with the detected luminosity $L=(2.2 
\pm 0.2)\times 10^{37}$ erg/s \citep{Mayer}. 

\begin{figure}
 \includegraphics[width=80mm]{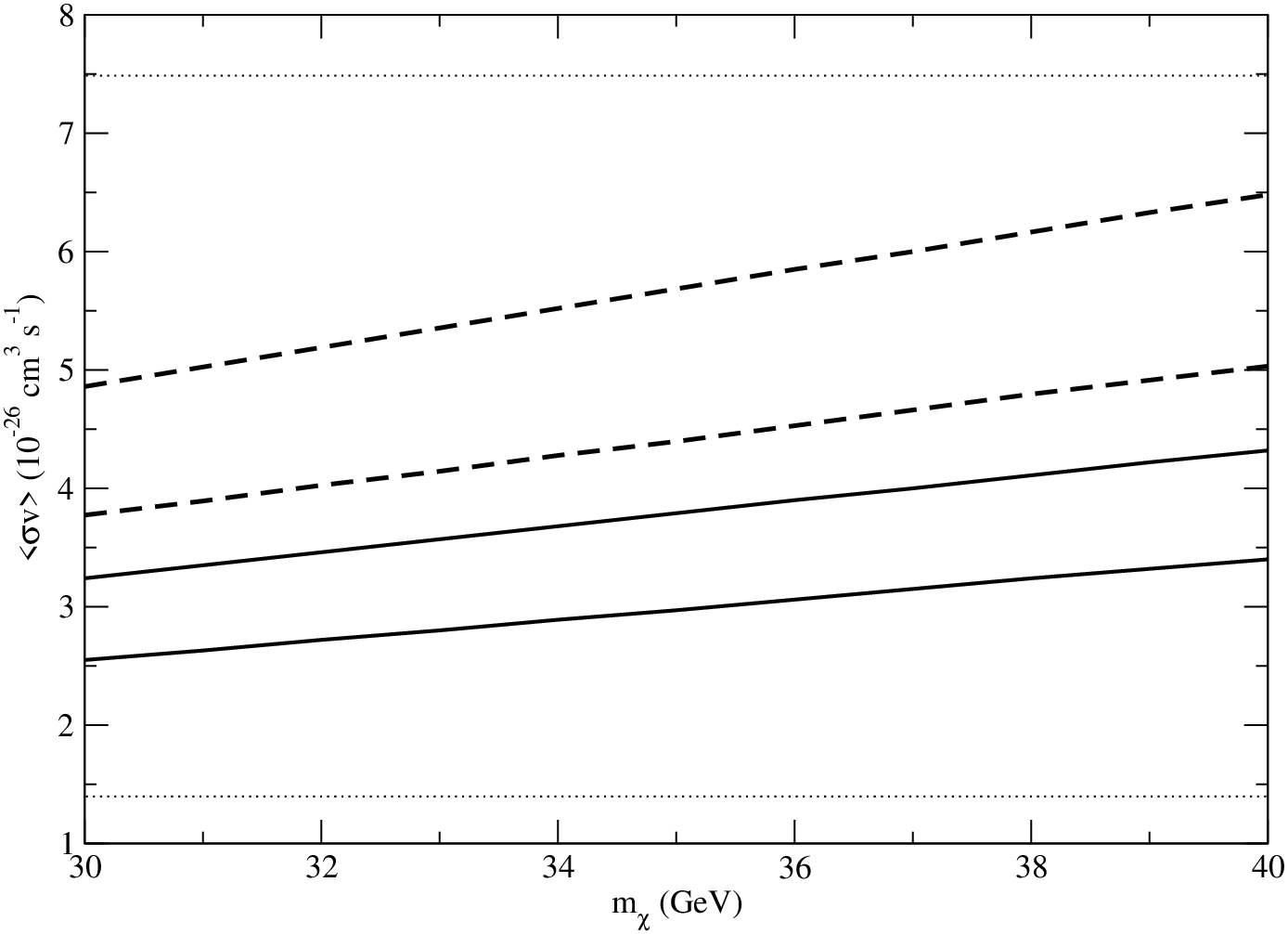}
 \caption{The parameter space constrained by the energy emitted from x-ray 
plasma is bounded by the solid lines (simple random walk diffusion model) 
and the dashed lines (turbulence model, assumed only 65\% of the energy is 
lost during the diffusion process). The dotted 
lines are the lower and upper limits of cross sections obtained from the 
GeV excess spectrum by using the Fermi-LAT data \citep{Abazajian,Daylan}.}
\end{figure}

\section{Discussion}
In this letter, we show that the annihilation of dark matter particles 
can satisfactorily explain the energy source of soft and hard components 
of hot plasma. The cross section 
and rest mass of dark matter particles calculated are consistent 
with the gamma ray observations and give excellent agreement 
with the prediction from the $b \bar{b}$ annihilation model based on the 
data from Fermi-LAT. The predicted cross section in our calculations 
($<\sigma v> \approx (2-6) \times 10^{-26}$ cm$^3$ s$^{-1}$) satisfies 
all the constraints from cosmic microwave background and low-redshift 
data \citep{Madhavacheril}, and also agrees 
with the canonical thermal relic abundance cross section in cosmology 
$<\sigma v> \approx 3 \times 10^{-26}$ cm$^3$ s$^{-1}$ \citep{Lopez}. 

In this model, the total number of positrons produced per one 
annihilation is $\sim 10$. For $r \le 20$ pc, we have $\Psi \sim 10^{38}$ 
s$^{-1}$. Since a positron takes $\sim 10^{13}-10^{14}$ s to loss all its 
energy, we 
predict that the total number of high energy positrons within 20 pc is 
about $10^{52}-10^{53}$ (the low energy positrons would annihilate with 
electrons to produce 511 keV emission line). Therefore, the number density 
of high energy positrons within 
20 pc is $n_{e^+} \sim 10^{-8}-10^{-7}$ cm$^{-3}$. Since the electron 
number density is $n_e \sim 0.1$ cm$^{-3}$ \citep{Muno}, the positron to 
electron 
ratio becomes $\sim 10^{-7}-10^{-6}$, which general agrees with the recent 
analysis in interstellar medium \citep{Vecchio}. Furthermore, this range 
of ratio can produce the observed 511 keV luminosity by the accretion of 
intergalactic material \citep{Vecchio}.

To conclude, we can notice that all the above evidences give a 
self-consistent picture in the Galactic Centre and strongly 
support the annihilation dark matter model. The rest mass and the cross 
section could probably be verified by the Large Hadron Collider Experiment 
in the future. 

\section{Acknowledgements}
I am grateful to the referee for helpful comments on the manuscript.

\label{lastpage}


\begin{thebibliography}{}
\bibitem[Abazajian et al. (2014)]{Abazajian} Abazajian, K. N., Canac, N., 
Horiuchi, S., and Kaplinghat, M., 2014, Phys. Rev. D, 90, 023526.
\bibitem[Belmont et al. (2005)]{Belmont} Belmont, R., Tagger, M., Muno, 
M., Morris, M., and Cowley, S., 2005, ApJ, 631, L53.
\bibitem[Bergstrom et al. (2005)]{Bergstrom} Bergstrom, L., Bringmann, T., 
Eriksson, M., and Gustafsson, M., 2005, Phys. Rev. Lett., 94, 131301.
\bibitem[Boehm et al. (2004)]{Boehm} Boehm, C., Hooper, D., Silk, J., 
Casse, M., and Paul, J., 2004, Phys. Rev. Lett., 92, 101301.
\bibitem[Borriello et al. (2009)]{Borriello} Borriello, E., Cuoco, A., and 
Miele, G., 2009, Phys. Rev. D, 79, 023518.
\bibitem[Calore et al. (2014)]{Calore} Calore, F., Cholis, I., McCabe, C., 
and Weniger, C., arXiv:1411.4647.
\bibitem[Cheng et al. (2006)]{Cheng} Cheng, K. S., Chernyshov, D. O., and 
Dogiel, V. A., 2006, ApJ, 645, 1138.
\bibitem[Cirelli et al. (2014)]{Cirelli} Cirelli, M., Gaggero, D., Giesen, 
G., Taoso, M., and Urbano, A., 2014, arXiv:1407.2173.
\bibitem[Crocker et al. (2010)]{Crocker} Crocker, R. M., Bell, N. F., 
Bal\'azs, C., and Jones, D., Phys. Rev. D, 81, 063516.
\bibitem[Daylan et al. (2014)]{Daylan} Daylan, T., Finkbeiner, D. P., 
Hooper, D., Linden, T. Portillo, S. K. N., Rodd, N. L., and Slatyer, T. 
R., 2014, arXiv:1402.6703.
\bibitem[Delahaye et al. (2008)]{Delahaye} Delahaye, T., Lineros, R., 
Donato, F., Fornengo, N., and Salati, P., 2008, Phys. Rev. D, 77, 063527.
\bibitem[Fornengo et al. (2012)]{Fornengo} Fornengo, N., Lineros, R. A., 
Regis, M., and Taoso, M., 2012, JCAP, 01(2012), 005.
\bibitem[Fritz et al. (2014)]{Fritz} Fritz, T. K. {\it et al.}, 2014, 
arXiv:1406.7568.
\bibitem[Gordon and Macias (2013)]{Gordon} Gordon, C., and Macias, O., 
2013, Phys. Rev. D, 88, 083521.
\bibitem[Hooper et al. (2013)]{Hooper} Hooper, D., Cholis, I., Linden, T. 
Siegal-Gaskins, J., and Slatyer, T., 2013, Phys. Rev. D, 88, 083009.
\bibitem[Izaguirre et al. (2014)]{Izaguirre} Izaguirre, E., Krnjaic, G., 
and Shuve, B., 2014, arXiv:1404.2018.
\bibitem[Lacroix et al. (2014)]{Lacroix} Lacroix, T., Boehm, C., and Silk, 
J., 2014, Phys. Rev. D, 90, 043508.
\bibitem[Longair (1994)]{Longair} Longair, M. S., {\it High Energy 
Astrophysics}, vol. 2 (Cambridge: Cambridge University Press, 1994).
\bibitem[Lopez-Honorez et al. (2013)]{Lopez} Lopez-Honorez, L., Mena, O., 
Palomares-Ruiz, S., and Vincent, A. C., 2013, JCAP, 07, 046.
\bibitem[Madhavacheril et al. (2014)]{Madhavacheril} Madhavacheril, M. S., 
Sehgal, N., and Slatyer, T. R., 2014, Phys. Rev. D, 89, 103508.
\bibitem[Mayer-Hasselwander et al. (1998)]{Mayer} Mayer-Hasselwander, H. 
A., {\it et al.}, 1998, Astron. Astrophys., 335, 161. 
\bibitem[Moore et al. (1999)]{Moore} Moore, B., Quinn, T., Governato, F., 
Stadel, J., and Lake, G., 1999, MNRAS, 310, 1147.
\bibitem[Muno et al. (2004)]{Muno} Muno, M. P., {\it et al.}, 2004, ApJ, 
613, 326.
\bibitem[Navarro et al. (1997)]{Navarro} Navarro, J. F., Frenk, C. S., 
White, S. D. M., 1997, ApJ, 490, 493.
\bibitem[Regis and Ullio (2008)]{Regis} Regis, M., and Ullio, P., 2008, 
Phys. Rev. D, 78, 043505.
\bibitem[Sakano et al. (2004)]{Sakano} Sakano, M. Warwick, R. S., 
Decourchelle, A., and Predehl, P., 2004, MNRAS, 350, 129.
\bibitem[Sidoli et al. (1999)]{Sidoli} Sidoli, L., Mereghetti, S., Israel, 
G. L., Chiapetti, L., Treves, A., and Orlanclini, M., 1999, ApJ, 525, 215.
\bibitem[Uchiyama et al. (2013)]{Uchiyama} Uchiyama, H., Hobukawa, M., 
Tsuru, T. G., and Koyama, K., 2013, PASJ, 65, 19.
\bibitem[Vecchio et al. (2013)]{Vecchio} Vecchio, A., Vincent, A. C., 
Miralda-Escude, J., and Pena-Garay, C., 2013, arXiv:1304.0324.
\bibitem[Wolfire et al. (1990)]{Wolfire} Wolfire, M. G., Tielens, A. G. G. 
M., and Hollenbach, D., 1990, ApJ, 358, 116.

\end{thebibliography}
\end{document}